\documentclass[prb,aps,twocolumn,showpacs,superscriptaddress]{revtex4-1}
\usepackage{amssymb}
\usepackage{stmaryrd}
\usepackage{hyperref}
\hypersetup{colorlinks=true, citecolor=blue, filecolor= black, linkcolor= blue, urlcolor= blue}
\usepackage{graphicx}
\usepackage{dcolumn}
\usepackage{bm}
\usepackage{float}
\begin{document}


\title{Two-Dimensional Magnetic Boron}

\author{Xiang-Feng Zhou}
\email{xfzhou@nankai.edu.cn}
\email{zxf888@163.com}
\affiliation{School of Physics and Key Laboratory of Weak-Light Nonlinear Photonics, Nankai University, Tianjin 300071, China}
\affiliation{Department of Geosciences, Center for Materials by Design, and Institute for Advanced Computational Science, Stony Brook University, Stony Brook, New York 11794, USA}
\author{Artem R. Oganov}
\affiliation{Department of Geosciences, Center for Materials by Design, and Institute for Advanced Computational Science, Stony Brook University, Stony Brook, New York 11794, USA}
\affiliation{Skolkovo Institute of Science and Technology, 5 Nobel St., Skolkovo 143025, Russia}
\affiliation{School of Materials Science, Northwestern Polytechnical University, Xi'an 710072, China}

\author{Zhenhai Wang}
\affiliation{Department of Geosciences, Center for Materials by Design, and Institute for Advanced Computational Science, Stony Brook University, Stony Brook, New York 11794, USA}
\affiliation{Peter Gr$\ddot{u}$nberg Research Center, Nanjing University of Posts and Telecommunications, Nanjing, Jiangsu 210003, China}

\author{Ivan A. Popov}
\affiliation{Department of Chemistry and Biochemistry, Utah State University, Logan, Utah 84322, USA}

\author{Alexander I. Boldyrev}
\affiliation{Department of Chemistry and Biochemistry, Utah State University, Logan, Utah 84322, USA}

\author{Hui-Tian Wang}
\affiliation{School of Physics and Key Laboratory of Weak-Light Nonlinear Photonics, Nankai University, Tianjin 300071, China}
\affiliation{National Laboratory of Solid State Microstructures and Collaborative Innovation Center of Advanced Microstructures, Nanjing University, Nanjing 210093, China}

\begin{abstract}
\noindent We predict a two-dimensional (2D) antiferromagnetic (AFM) boron (designated as $M$-boron) by using \textit{ab initio} evolutionary methodology. $M$-boron is entirely composed of B$_{20}$ clusters in a hexagonal arrangement. Most strikingly, the highest valence band of $M$-boron is isolated, strongly localized, and quite flat, which induces spin polarization on each cap of the B$_{20}$ cluster. This flat band originates from the unpaired electrons of the capping atoms, and is responsible for magnetism. $M$-boron is thermodynamically metastable and is the first cluster-based 2D magnetic material in the elemental boron system.
\end{abstract}



\maketitle
\section{INTRODUCTION}
Boron, a nearest neighbor of carbon in the Periodic Table, is an element of unique chemical and structural complexity. Its electron-deficient bonding leads to rich diversity of its crystal structures.\cite{R01,R02} Boron is in many ways an analog of carbon, its nanostructures--clusters, fullerenes, nanotubes, and 2D structures had attracted enormous attention,\cite{R03,R04,R05,R06,R07,R08,R09,R10,R11,R12,R13,R14,R15,R16,R17,R18} in the hopes of discovering unique properties, surpassing carbon. A recent prediction of the B$_{80}$ buckyball has aroused extensive interest because of structural similarity to the C$_{60}$ fullerene.\cite{R05,R19} It was followed immediately by the proposal of stable 2D boron sheets with triangular and hexagonal motifs (named as $\alpha$-sheet),\cite{R06} which can serve as a building block (or precursor) for fullerenes, nanotubes, and nanoribbons. However, subsequently, buckled bilayer structures were predicted to be much more stable. Some of them turned out to have novel electronic properties, such as distorted Dirac cones.\cite{R16} For carbon, magnetism was reported in the highly oriented pyrolitic graphite with vacancies,\cite{R20,R21} graphite ribbons under electric field,\cite{R22} tetrakis(dimethylamino)ethylene fullerene, and so on.\cite{R23} Unlike carbon, elemental boron has not been frequently reported to possess magnetism,\cite{R24} much less for 2D boron even though it is a graphene analogue.\cite{R25,R26} Most recently, boron sheets were grown on the Ag $(111)$ surface under pristine ultra-high vacuum conditions.\cite{R27} To resolve the crystal structures of several samples, we focused on some low-energy 2D structures with hexagonal symmetry, which should match the lattice of Ag $(111)$ substrate very well. This resulted in a serendipitous discovery of $M$-boron.
\section{METHOD}
The evolutionary structure searches were conducted with 18, 20, 22, 24, 26, and 28 atoms per unit cell using the \textsc{uspex} code.\cite{R28,R29} The prime intention is to search for some thermodynamically favored, thickness-depended 2D boron structures, especially for some 2D structures, which contain B$_{12}$ icosahedra (its thickness is $\sim$3.7~{\AA}).\cite{R17} Therefore, the initial thickness for 2D boron structures was set to 4~{\AA} and allowed to change during relaxation. There are many low-energy structures found by \textsc{uspex}, only one example (20 atoms per cell) was presented for discussion due to the space limit. The final structure relaxations used the all-electron-projector augmented wave method as implemented in the Vienna \textit{ab initio} simulation package \textsc{vasp}.\cite{R30,R31} The exchange-correlation energy was computed within the generalized gradient approximation (GGA) with the functional of Perdew, Burke, and Ernzerhof (PBE).\cite{R32} The plane wave  cutoff energy of 600 eV and the uniform $\Gamma$-centered $k$-points grids with resolution of $2 \pi \times 0.04$~\AA$^{-1}$ were used. In addition, the hybrid PBE0 functional was also employed to confirm the energetic stability of several 2D boron structures.\cite{R33} Phonon dispersion curves were computed using the \textsc{phonopy} package with the $3 \times 3 \times 1$  supercell for nonmagnetic (NM), ferromagnetic (FM), and AFM $M$-boron.\cite{R34} Spin-polarized density functional theory calculations were adopted for investigations of magnetism. For the Adaptive Natural Density Partitioning ($AdNDP$) chemical bonding analysis we chose an appreciably large fragment consisting of seven B$_{20}$ fragments,\cite{R35} which a priori should contain all the bonding elements of the whole 2D system due to its high symmetry and appropriate size. PBE0/6-31G level of theory was used. It was previously shown that $AdNDP$ is not sensitive to the level of theory or the basis set.\cite{R36} All the $AdNDP$ calculations were performed using the Gaussian 09 software package.\cite{R37} Molecular visualization was performed using Molekel 5.4.0.8.\cite{R38}

\begin{figure}[h]
\begin{center}
\includegraphics[width=8cm]{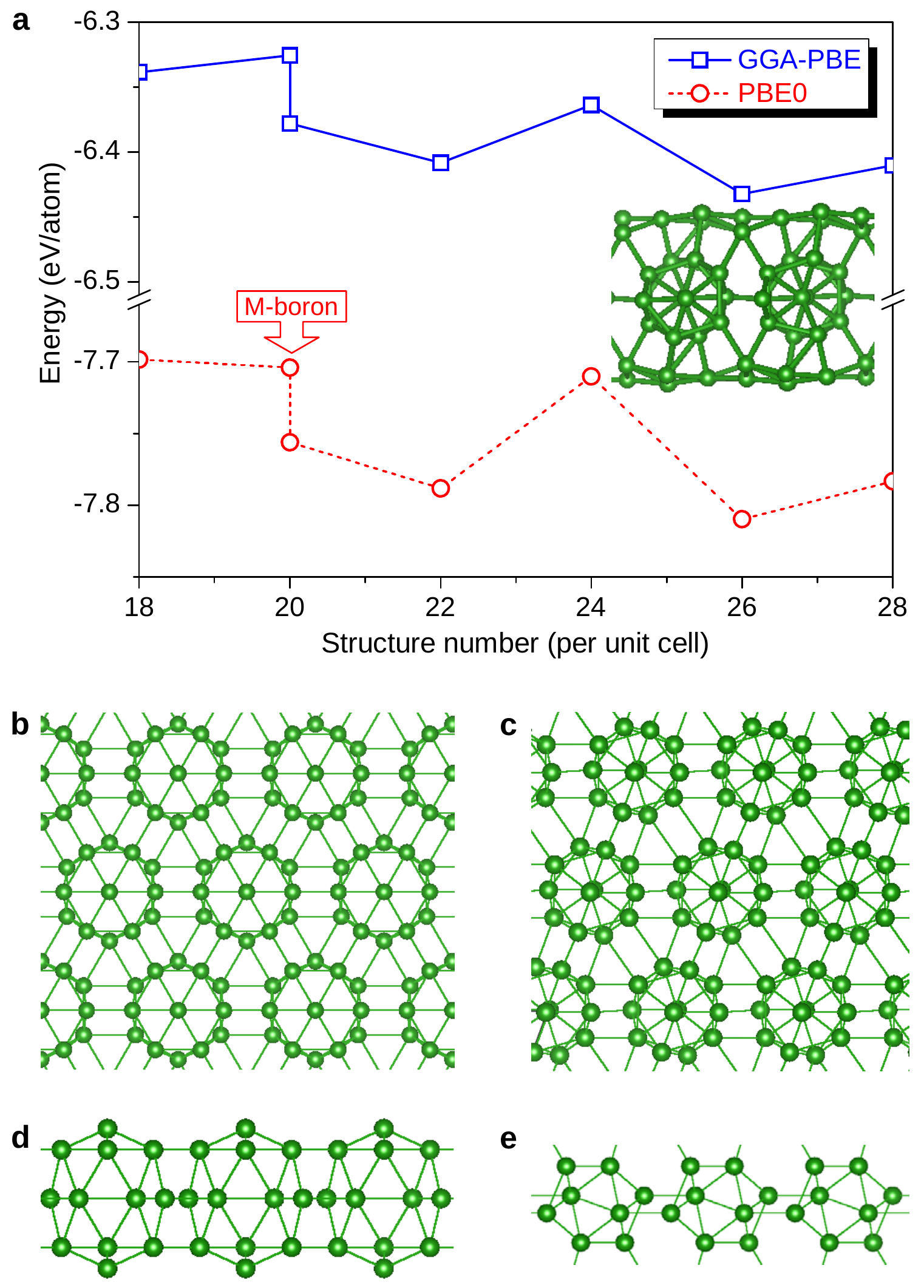}
\caption{%
(Color online) (a)	The total energies from the GGA-PBE and PBE0 calculations for several 2D boron structures. The inset shows the structure of 2D-B$_{26}$. (b) and (d) projection of $M$-boron along the $[001]$ and $[100]$ directions. (c) and (e) projection of $\alpha$-boron along the $[111]$ and $[1 1 \bar{2}]$ directions.}
\end{center}
\end{figure}
\section{RESULTS AND DISCUSSION}
The most stable structures from separated structural searches are designated from 2D-B$_{18}$, 2D-B$_{20}$, ..., 2D-B$_{28}$. As shown in Fig.~1(a), the GGA-PBE (PBE0) results yield the total energies for 2D-B$_{18}$, 2D-B$_{20}$, 2D-B$_{22}$, 2D-B$_{24}$, 2D-B$_{26}$, and 2D-B$_{28}$ are -6.338 (-7.698), -6.378 (-7.756), -6.408 (-7.788), -6.364 (-7.710), -6.432(-7.81), and -6.41 (-7.783) eV/atom, respectively. Therefore, GGA-PBE and PBE0 give the same ranking of structures by energetic stability. The inset of Fig.~1(a) shows the structure of 2D-B$_{26}$, where the icosahedral B$_{12}$ units are covered by two monolayer sheets (top and down). 2D-B$_{26}$ has the lowest energy among all of 2D boron allotropes. Its $P\bar{1}$ symmetry indicates that multilayered 2D boron prefers to be a frustrated system.\cite{R16} The electronic structure calculations show that 2D-B$_{22}$ and 2D-B$_{28}$ are semiconductors with the DFT bandgap of 0.45 and 0.51 eV. In contrast, other stable 2D boron forms are metals. The GGA-PBE and PBE0 results for $M$-boron are -6.326 and -7.704 ev/atom. This means that $M$-boron is a metastable phase. The crystal structures of $M$-boron and bulk $\alpha$-boron are shown in Figs.~1(b)--1(e).\cite{R39} The $(001)$ projection of $M$-boron (Fig.~1b) shows dodecagonal patterns, which look very similar to those of $(111)$ plane of $\alpha$-boron (Fig.~1c). The difference is that the latter forms decagonal motif. $M$-boron is completely made of B$_{20}$ cages, in contrast to the well-known B$_{12}$ icosahedra in $\alpha$-boron (Fig.~1d, 1e). It has three layers: the first and third layers are composed of hexagonal pyramids with mirror symmetry, separated by the second flat layer. Table I lists lattice constants, atomic positions, and total energies of NM, FM, and AFM states of $M$-boron. It shows that the AFM ordering is the ground state of $M$-boron, which has almost the same energy (-6.33 eV/atom) as the recently predicted 2D Dirac boron, and lower in energy than $\alpha$-sheet (-6.28 eV/atom).\cite{R16} Because there is only one B$_{20}$ cluster per unit cell of $M$-boron, to confirm its magnetic surface state, we had to build a $2 \times 2 \times 1$ supercell with FM and AFM orderings for the surface, respectively. The calculations show that the FM surface state (-126.513 eV/cell) is more stable than the AFM suface state (-126.509 eV/cell). This indicates that $M$-boron is a 2D AFM material associated with the FM surface state. Interestingly, size-selected boron clusters have been found to be 2D up to at least 25 atoms as anions in gas-phase experiments.\cite{R40,R41,R42,R43} Even though a neutral B$_{20}$ cluster was computationally suggested to have a 3D double ring structure,\cite{R44} an infrared/ultraviolet double ionization experiment failed to detect this structure.\cite{R45} Importantly, the 20-atom boron cluster constituting the infinitely extended $M$-boron has a different structure than the proposed tubular structure of B$_{20}$.\cite{R40} Combined theoretical and experimental studies of gas-phase anionic clusters were comprehensively reviewed elsewhere.\cite{R46,R47}

\begin{figure}[h]
\begin{center}
\includegraphics[width=8cm]{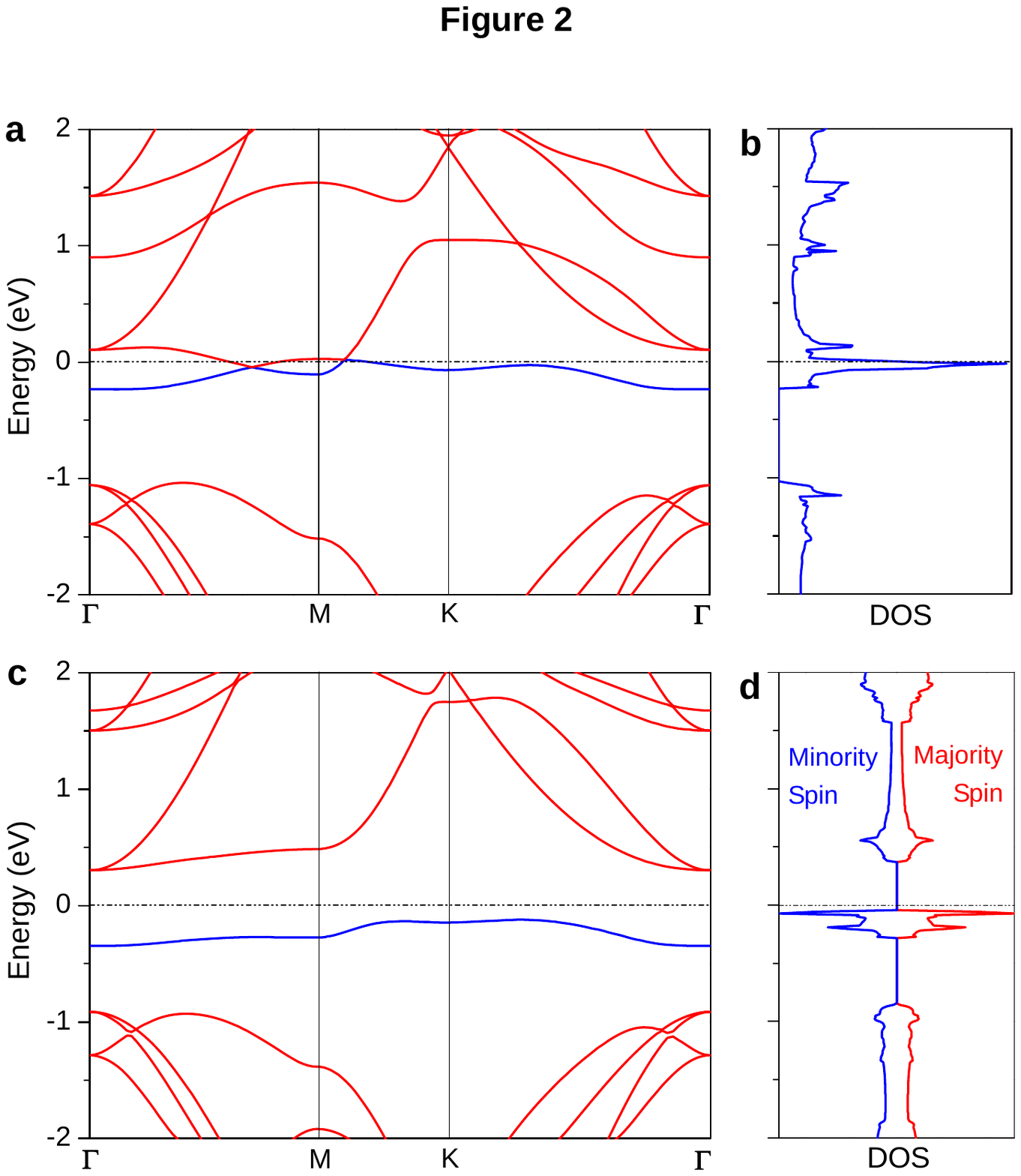}
\caption{%
(Color online) Electronic properties of $M$-boron. (a) Band structure of NM $M$-boron. The Fermi level is set to zero. Most of the flat band is colored in blue hollow circles; part of it is colored in red within the band crossing near the M point, (b) Density of states (DOS) of NM $M$-boron, (c) Band structure of AFM $M$-boron. The highest valence band (flat band) is colored in blue, (d) DOS of AFM $M$-boron. The majority spin is colored in red, and blue for the minority spin.}
\end{center}
\end{figure}

\begin{table}
\caption{Lattice constants, atomic positions and total energies ($E_{tot}$) of $M$-boron with NM, FM, and AFM states.}
\begin{tabular}{lllllc}
\hline\hline
Phase & $a$    & $b$    & $c$    & Atomic positions & $E_{tot}$ \\
      &({\AA}) &({\AA}) &({\AA}) &                  & (eV/atom) \\
\hline
NM  & 5.178 & 5.178 & 16.581 & B$_{1}$ (0.0, 0.0, 0.366)   & -6.312 \\
    &       &       &        & B$_{2}$ (0.668, 0.0, 0.596) &  \\
    &       &       &        & B$_{3}$ (0.793, 0.586, 0.5) &  \\
FM  & 5.194 & 5.194 & 16.451 & B$_{1}$ (0.0, 0.0, 0.365)   & -6.321 \\
    &       &       &        & B$_{2}$ (0.666, 0.0, 0.596) &  \\
    &       &       &        & B$_{3}$ (0.792, 0.584, 0.5) &  \\
AFM & 5.189 & 5.189 & 16.534 & B$_{1}$ (0.0, 0.0, 0.363)   & -6.326 \\
    &       &       &        & B$_{2}$ (0.667, 0.0, 0.595) &  \\
    &       &       &        & B$_{3}$ (0.793, 0.586, 0.5) &  \\
\hline\hline
\end{tabular}
\end{table}

\begin{figure}[h]
\begin{center}
\includegraphics[width=8cm]{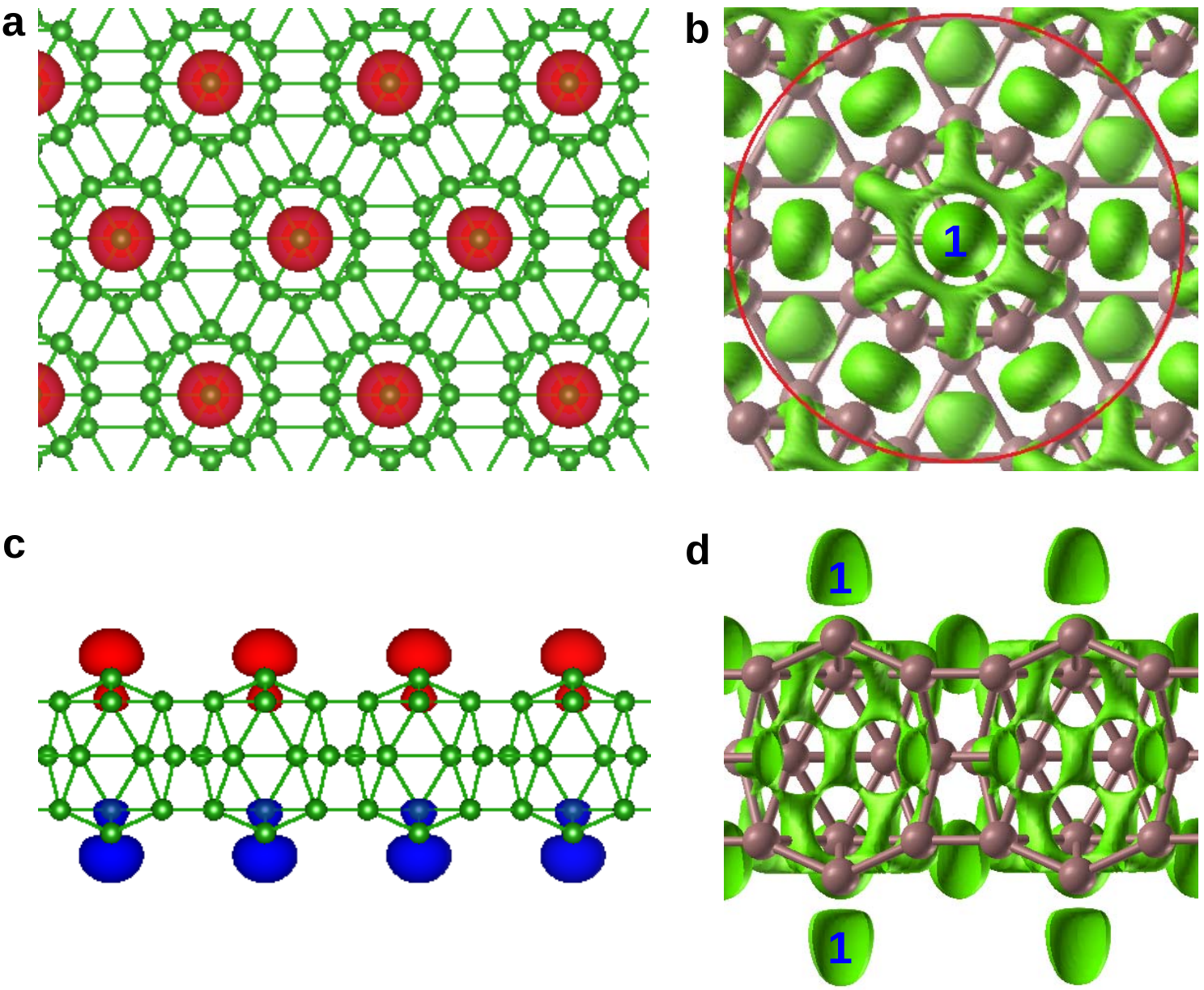}
\caption{%
(Color online) Band-decomposed spin charge density and electron localization function (ELF) of AFM $M$-boron: (a) and (c) projection of the flat band decomposed majority spin charge density (colored in red) along the $[001]$ and the minority spin charge density (colored in blue) along the $[100]$ directions; (b) and (d) projection of the ELF in $M$-boron along the $[001]$ and $[100]$ directions. The unpaired electrons are labeled as number 1 in $b$ and $d$.}
\end{center}
\end{figure}

NM $M$-boron is metallic with a flat band (the highest valence band) at the Fermi level,  and the Fermi level crosses both the lowest conduction band and the flat band around the $M$ point, leading to large density of states (DOS) at the Fermi level (Fig.~2a, 2b). To stabilize the material, the high DOS at the Fermi level can be lowered either by structural (Peierls) distortion or by spin polarization.\cite{R48} Since the crystal structure of $M$-boron is highly symmetric and strong B-B bonds make distortions difficult, the flat band may induce spin polarization of $M$-boron according to the Fermi level instability. The band structure and DOS of AFM $M$-boron are shown in Fig.~2(c) and 2(d).  Due to spin polarization, the flat band is separated away from the highly dispersive conduction band, resulting in a NM-AFM transition and opening of a bandgap; $M$-boron is an AFM semiconductor with the indirect bandgap of 0.43 eV (Fig.~2c). Actually, as mentioned above, for $M$-boron we see order of stability NM $<$ FM $<$ AFM, which is in good agreement with the electronic stability among different states of $M$-boron (NM metal $<$ FM metal\cite{R49} (Fig.~S1) $<$ AFM semiconductor). Most importantly, the isolated flat band with a dispersion of 0.23 eV (Fig.~2c) may play a decisive role in the emergence of magnetism. To explore the physical origin of the flat band, the band decomposed spin-polarized charge density at all $k$ points is plotted in Fig.~3(a) and 3(c). Large bubbles of spin density are localized on the top of B$_{1}$ atoms (the polar sites of the hexagonal pyramids). These bubbles represent the majority (colored in red) and minority (colored in blue) spin electrons and have the same size and shape due to the mirror symmetry. The opposite spin densities above and below the plane show that $M$-boron is an AFM semiconductor associated with the FM surface state. The charge density distribution for the flat band is predominantly derived from the out-of-plane ($p_z$ orbitals) states of B$_{1}$ atoms. This implies that the strongly localized flat band, mostly originating from the $p_z$ orbitals of the B$_{1}$ atoms, is responsible for magnetism of $M$-boron.

\begin{figure*}
\begin{center}
\includegraphics[width=1.2\columnwidth]{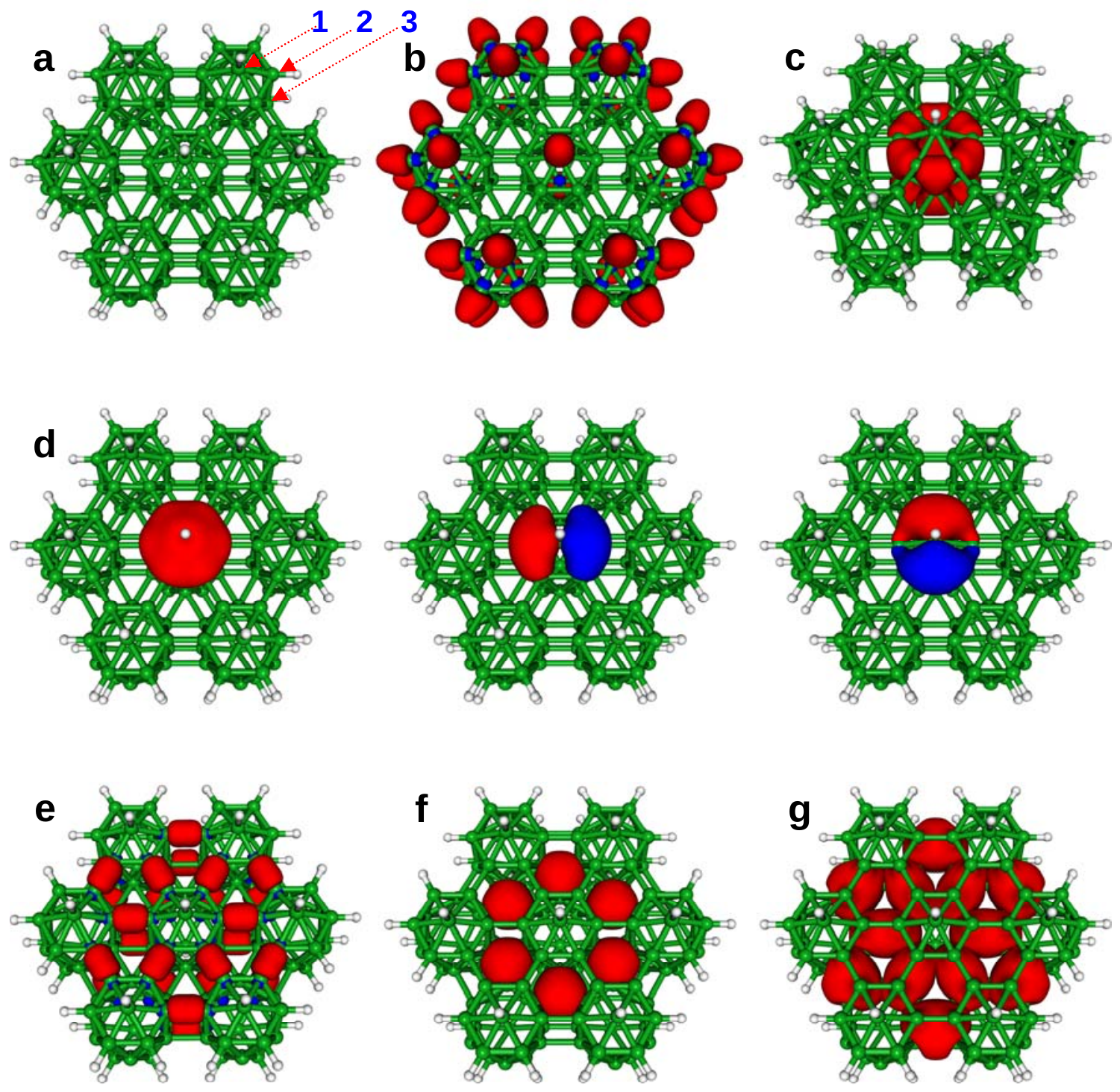}
\end{center}
\caption{\label{fig4}
(color online) Chemical bonding analysis for the model fragment of B$_{20}$ clusters in $M$-boron: (a) B$_{140}$H$_{50}$$^{34+}$, where three inequivalent boron atomic positions are labeled as B$_1$, B$_2$, and B$_3$; (b) fifty $2c-2e$ B-H $\sigma$ bonds with occupation numbers (ONs) in the 1.50-1.70 $|e|$ range, (c) twelve $3c-2e$ B$_2$-B$_3$-B$_2$ $\sigma$ bonds with ONs = 1.82 $|e|$ forming the side frame of each B$_{20}$ cluster, (d) three $7c-2e$ $\sigma$ bonds with ONs = 1.84-1.93 $|e|$ found on every cap. Intercluster bonding elements: (e) twenty four classical $2c-2e$ B$_2$-B$_2$ $\sigma$ bonds with ONs = 1.81 $|e|$ responsible for the direct bonding between B$_{20}$ building blocks, (f) six $3c-2e$ B$_3$-B$_3$-B$_3$ $\sigma$ bonds with ONs = 1.95 $|e|$ found in the middle layer, (g) twelve $8c-2e$ $\sigma$ bonds ONs = 1.96 $|e|$ connecting all three layers of the $M$-boron. All the bonding elements of each type are superimposed on a single framework.}
\end{figure*}

The B$_{12}$ cluster in $\alpha$-boron has 36 valence electrons, 26 of which may be used for intraicosahedral bonds and 10 for intericosahedral bonds. Each icosahedron forms six two-center-two-electron ($2c-2e$) bonds with the icosahedra of neighboring layers, which requires $6 \times 2/2 = 6$ electrons, as well as six closed three-center-two-electron ($3c-2e$) bonds with the neighboring icosahedra in its own layer, these multicenter bonds require $6 \times 2/3 = 4$ electrons. Therefore, the electrons are balanced for $\alpha$-boron according to Wade¡¯s rule.\cite{R01} Since $M$-boron is also a semiconductor and made entirely of B$_{20}$ clusters, it is natural to investigate the relationship between bonding and semiconducting nature of this phase. As listed in Table I, there are three inequivalent atomic positions (B$_{1}$, B$_{2}$ and B$_{3}$). For the $AdNDP$ chemical bonding analysis, an appreciably large fragment consisting of seven B$_{20}$ fragments was chosen, which a priori should contain all the bonding elements of the whole 2D system due to its high symmetry and appropriate size (Fig.~4a). Since M-boron is found to be AFM in its ground state, we intentionally quenched all fourteen the B$_{1}$ atoms with H atoms to deal with a single-configuration wave-function. It is important to note that this inclusion does not affect a chemical bonding pattern of the 2D system, and, presumably, should result in the formation of fourteen B$_{1}$-H bonds. Next, 36 additional H atoms were added to terminate the peripheral bonding interactions of six outer B$_{20}$ fragments with the rest of $M$-boron. Clearly, number of electrons in the proposed cluster (B$_{140}$H$_{50}$), which serves as a model fragment of the infinitely extended 2D $M$-boron, should be adjusted for the boundary conditions between the cluster being inside of the solid state system and its neighboring atoms, with which it shares electrons. In order to account for those shared electrons between the cluster and the neighboring atoms, the external positive charge of +34 was set. Thus, the model cluster considered for the chemical bonding analysis is B$_{140}$H$_{50}$$^{34+}$. As shown in Fig.~4(b), the $2c-2e$ B$_{2}$-H and B$_{1}$-H bonds found in the B$_{140}$H$_{50}$$^{34+}$ model correspond to B$_{2}$-B$_{2}$ bonds and the lone $p_z$ electrons on B$_1$ atoms in the periodic structure of the $M$-boron, respectively. The $AdNDP$ chemical bonding analysis revealed two types of bonding. For intracluster bonding, there are twelve $3c-2e$ bonds (similar to the case of all-boron fullerene\cite{R18}) (Fig.~4c) on the sides and six seven-center-two-electron ($7c-2e$) bonds (Fig.~4d) on each cap of the B$_{20}$ cluster, which requires $12 \times 2+6 \times 2 = 36$ electrons. For intercluster bonding, there are twelve $2c-2e$ bonds (Fig.~4e) which connect the neighboring B$_{20}$ clusters with B$_{2}$ atoms (at top layer), six $3c-2e$ bonds (Fig.~4f) which connect the B$_{20}$ clusters with B$_{3}$ atoms (at middle layer), and six eight-center-two-electron ($8c-2e$) bonds (Fig.~4g) formed within all three layers of $M$-boron (each cluster contributes one electron). This requires $12 \times 2/2+6 \times 2/3+6 \times 1 = 22$ electrons. Occupation numbers for different kinds of bonds are close to the ideal value of 2 electrons, hence giving additional credibility to the chemical bonding picture of $M$-boron. The most striking feature, there are two localized unpaired electrons (labeled as 1 in Fig.~3b and 3d) on the B$_{1}$ atoms are the key factors for the existence of magnetism. In total, there are $36+22+2 = 60$ electrons, which show that $M$-boron satisfies the electron counting rule,\cite{R50,R51} becoming an AFM semiconductor. For FM $M$-boron, the magnetic moment is 1.71 $\mu_{B}$/cell. Because B$_{1}$ atoms govern magnetism, the local magnetic moment for B$_{1}$ atoms should be 0.85 $\mu_{B}$ (~one electron per atom). Given that AFM $M$-boron has a FM surface state, we can reasonably infer that B$_{1}$ atoms in the AFM state may have the same local magnetic moment of ~1 $\mu_{B}$.

\begin{figure}[h]
\begin{center}
\includegraphics[width=8cm]{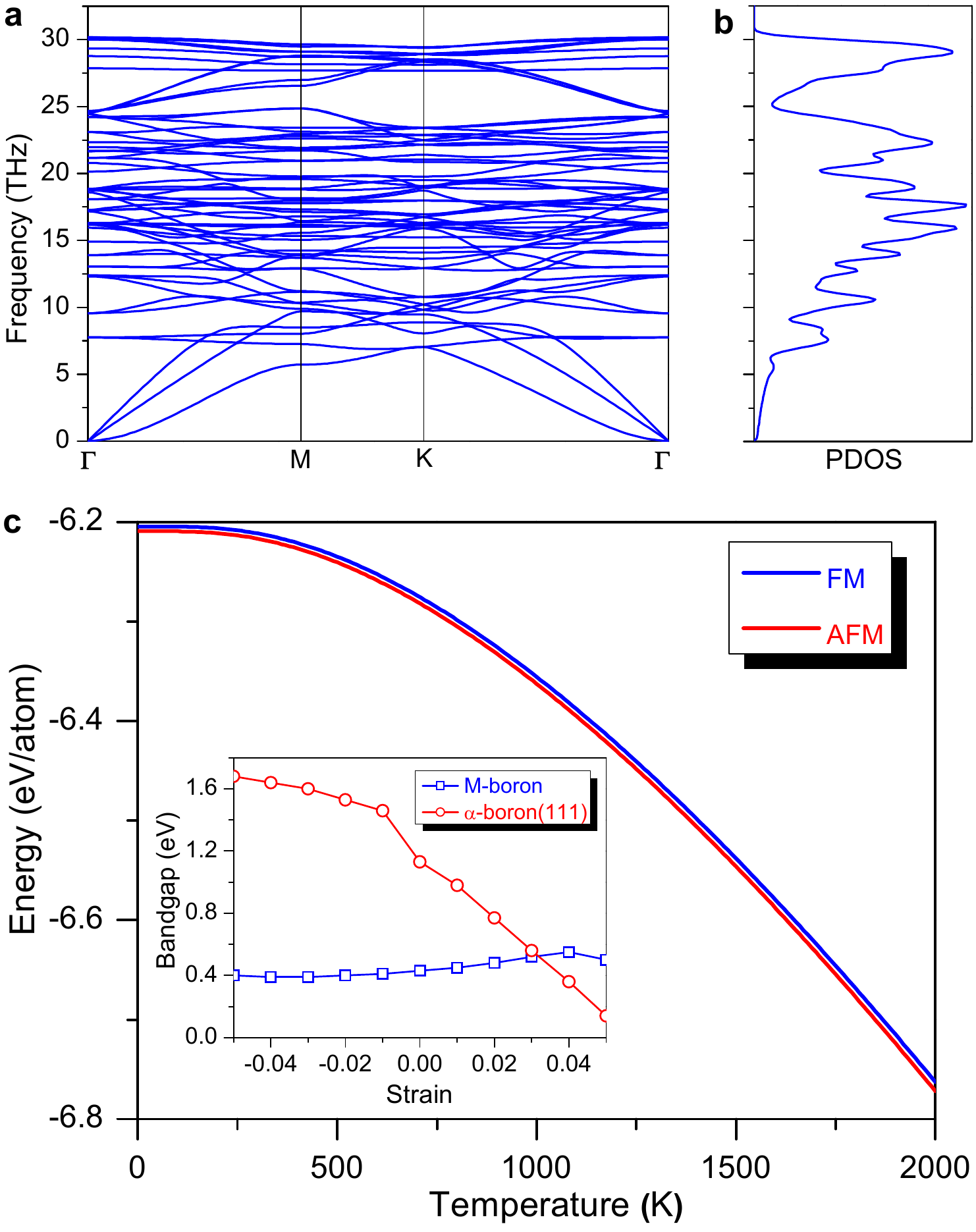}
\caption{%
(Color online) Stability of $M$-boron. (a) Phonon dispersion curves of AFM $M$-boron, (b) Phonon density of states (PDOS) of $M$-boron, (c) Temperature dependence of the free energy for the FM and AFM states. The inset shows variation of bandgap of $M$-boron compared with $\alpha$-boron $(111)$ surface via the biaxial tensile (negative strain) or compressive strain (positive strain).}
\end{center}
\end{figure}

Phonon dispersion curves and phonon density of states (PDOS) show that both FM (Fig.~S2a) and AFM (Fig.~5a and 5b) versions of $M$-boron are dynamically stable, but there are large imaginary frequencies in the NM state (Fig.~S2b). \cite{R49} Usually, structural phase transitions are expected to remove soft-mode instabilities. However, for $M$-boron, the electronic topological transition (the emergence of magnetism) is verified to be an alternative way to resolve the dynamical instability. Moreover, in order to take the zero point energy and temperature effect into account, we computed harmonic free energy for both the FM and AFM states. The temperature dependence of the free energy for the FM and AFM $M$-boron is shown in Fig.~5(c). It shows that AFM $M$-boron is always more stable than FM $M$-boron up to the temperature of 2000 $K$. Additionally, since the electronic properties of materials can be tuned by external strain, it is of great interest to study the influence of biaxial tensile or compressive strain on the bandgap of AFM $M$-boron with respect to the reconstructed $\alpha$-boron $(111)$ surface (Fig.~5c).\cite{R17} The bandgap of the $\alpha$-boron $(111)$ surface increases or decreases with biaxial tensile or compressive strain, respectively. The semiconducting $\alpha$-boron $(111)$ surface shows a normal strain behavior just like the compression behavior of $\alpha$-boron under pressure.\cite{R02} In contrast, the bandgap of $M$-boron is almost insensitive to strains of up to 5\%. The position and width of the flat band are almost fixed and independent of tensile or compressive strain, guaranteeing robustness of magnetism. $M$-boron can endure large strain, persisting in its AFM state. Therefore, it is practicable to grow $M$-boron on some metal surfaces and expect it to be magnetic,  because the lattice mismatch is less than 4\% for Ag $(111)$, 2\% for Cu $(111)$.

\section{CONCLUSION}
In summary, we have predicted 2D magnetic boron by using an \textit{ab initio} evolutionary structure search combined with spin-resolved electronic structure calculations. $M$-boron bridges the gap between boron clusters and 2D nanomaterials, which opens a new path to explore light elemental magnetic materials at atomic scale.

\textbf{ACKNOWLEDGMENTS} X.F.Z thanks Bogeng Wang and Xiangang Wan for valuable discussions. This work was supported by the National Science Foundation of China (Grant No. 11174152), the National 973 Program of China (Grant No. 2012CB921900), the Program for New Century Excellent Talents in University (Grant No. NCET-12-0278), and the Fundamental Research Funds for the Central Universities (Grant No. 65121009). A.R.O. thanks the National Science Foundation (Grants No. EAR-1114313, No. DMR-1231586), DARPA (Grant No. W31P4Q1210008), DOE (Computational Materials and Chemical Sciences Network Project No. DE-AC02-98CH10886), the Government of Russian Federation (Grant No. 14.A12.31.0003), and the Foreign Talents Introduction and Academic Exchange Program (Grant No. B08040). Part of this work was partly performed on the cluster of the Center for Functional Nanomaterials, Brookhaven National Laboratory, which is supported by the DOE-BES under Contract No. DE-AC02-98CH10086. Z.W. thanks the China Scholarship Council (No. 201408320093) and the Natural Science Foundation of Jiangsu Province (Grant No. BK20130859). This work was also supported by the National Science Foundation (CHE-1361413 to A.I.B.). Computer, storage, and other resources from the Division of Research Computing in the Office of Research and Graduate Studies at Utah State University are gratefully acknowledged.



\end{document}